\documentstyle[12pt]{article}
\renewcommand{\b}{\beta}

\newcommand{\D}{\Delta}

\newcommand{\la}{\lambda}
\renewcommand{\a}{\alpha}
\begin{document}
\title{Quantum Computer with Fixed Interaction is Universal}
\author{Yuri Ozhigov\thanks{e-mail: 
ozhigov@ftian.oivta.ru.}\ \ \ and \ \ Leonid Fedichkin\thanks{e-mail: 
leonid@ftian.oivta.ru.} \\[7mm]
Institute of Physics and Technology,\\
Russian Academy of Sciences,\\ 
Nakhimovsky pr. 34,\\ 
Moscow, 117218, Russia}
\date{}
\maketitle
\begin{abstract} It is proved that a quantum computer with fixed and 
permanent interaction of diagonal type between qubits 
proposed in the work quant-ph/0201132 is universal. 
Such computer is controlled only by one-qubit quick transformations, 
and this makes it feasible. 
\end{abstract}

\section{Introduction and background}

A model of quantum computer with fixed and permanent interaction 
between qubits was proposed in the paper \cite{Oz} 
where it was shown how to implement QFT and simulation of 
wave functions dynamics by such computer. 
In this paper we prove that such a model is universal 
that is every quantum algorithm can be implemented in the framework 
of this model with only linear slowdown for long-distance 
interaction and with the slowdown as multiplication by a size of memory 
for short distance interaction. 
Here we have to suppress undesirable interactions like it is done 
in the work \cite{Fe}. 
But now we shall use the method of random strings consisting of NOT 
operations proposed in \cite{Oz} which uses a diagonal form of interactions. 
Surprisingly, that our method of suppressing undesirable interactions 
does not depend on individual features of qubits. 

A traditional way for implementation of quantum algorithms 
requires a control on two qubits level that is an ability 
to "switch on" and to "switch off" interaction between qubits. 
Whereas is widely known that two qubits transformations is a stumbling 
block in quantum computing in view of technological difficulties. 
The most natural way is to use a fixed and permanent interaction 
between qubits and control a process of computation by only 
one-qubit transformations. This way gives a universal quantum computer 
if our fixed interaction has a diagonal type. 
Note that it is no matter how a fixed interaction decreases depending 
on the distance between qubits, for example it may be nonzero only 
for the neighboring qubits, etc. 

A permanent interaction between qubits in our computer depends 
only on their spatial disposition which is fixed. 
The only condition we impose to the interaction is that it must be diagonal. 
Thus if $j$ and $k$ denote identification numbers of two qubits then the 
Hamiltonian of their interaction will have the form
\begin{equation}
{\rm A)}\  H_{j,k}=\left(
\begin{array}{ccccc}
&E^{j,k}_1 &0 &0 &0\\
&0 &E^{j,k}_2 &0 &0\\
&0 &0 &E^{j,k}_3 &0\\
&0 &0 &0 &E^{j,k}_4
\end{array}
\right) ,
\ \ \ \ \ \ \ \ \ \ \ \ \ 
{\rm B)}\ H_{j,k}=\left(
\begin{array}{ccccc}
&0 &0 &0 &0\\
&0 &0 &0 &0\\
&0 &0 &0 &0\\
&0 &0 &0 &E_{j,k}
\end{array}
\right)
,\ \ E_{j,k} >0. 
\label{Ham} 
\end{equation}

At first note that any interaction of the general form (\ref{Ham}, A) 
may be reduced to the form (\ref{Ham}, B) by adding appropriate 
one-qubit Hamiltonians $H'_{j,k}$ which matrices have the forms
$$
\left(
\begin{array}{ccccc}
&a &0 &0 &0\\
&0 &a &0 &0\\
&0 &0 &b &0\\
&0 &0 &0 &b\\
\end{array}
\right),
\ \ \ \ \ \ \ \ \ 
\left(
\begin{array}{ccccc}
&\a &0 &0 &0\\
&0 &\b &0 &0\\
&0 &0 &\a &0\\
&0 &0 &0 &\b\\
\end{array}
\right)
.
$$
This addition reduces Hamiltonian of the form (\ref{Ham}, A) to (\ref{Ham}, B) 
and it can be alternatively fulfilled by one-qubit quick transformations because 
all these diagonal matrices commute. 

Note that the different pairs of qubits may interact variously, 
they may be disposed with the different intervals and be not placed along one 
line, etc. 

\section{Suppression of undesirable interactions by one qubit operations}

To prove a universality of computational model we must show how 
one can fulfil an arbitrary two qubits operation. 
Given a unitary transform induced by Hamiltonian (\ref{Ham}, B) 
in time frame 1: $U_{j,k}=\exp (-iH_{j,k})$ (Plank constant equals 1). 
In fact it would suffice to fulfil this transform on two qubits: $j$th 
and $k$th preserving all others untouched. Just this last condition 
is difficult to guarantee for permanent interaction. If we can do it 
when at first we can fulfil an arbitrary two qubits operation with 
every separated pair of qubits. Then for a long distance interaction 
we shall have at most a linear slowdown, for a short distance interaction 
we shall need to perform SWAP operations to bring a required pair of 
qubits together and thus obtain time factor equal to a size of memory. 

Now show how to implement $U_{j,k}$. If we simply wait for a time 1, 
when we obtain a transformation $U_{j,k}\bigotimes U'\bigotimes\ldots\bigotimes U''$ where all $U'$ have the form $U_{j',k'}$ where $\{ j',k'\}\neq \{j,k\}$. We should get rid of these interactions. 
Declare $j$th and $k$th qubits separated. 

We shall apply one qubit gate NOT several times to all qubits 
but separated ones to suppress all two qubits interactions excluding 
interaction between separated qubits. 
For each not separated qubit number $p$ consider the Poisson 
random process ${\cal A}_p$ generating time 
instants $0<t_1^p <t_2^p <\ldots <t_{m_p}^p <1$ with some fixed 
density $\la\gg 1$. Let all ${\cal A}_p$ are independent. 
Now fulfil transformations NOT on each $p$th qubit in 
instants $t^p_m$ sequentially. In instant 1 fulfil NOT on $p$th 
qubit if and only if $m_p$ is odd. Then after this procedure each 
$q$th qubit restores its initial value $a_q$. Count the phase shift 
generated by this procedure. Interaction between separated qubits 
remains unchanged. Fix some not separated qubit number $p$ and count 
its deposit to phase. It consists of two summands: the first comes 
from interaction with separated and the second - from interaction 
with not separated qubits. Count them sequentially.

1. In view of big density $\la$ of Poisson process ${\cal A}_p$ 
about half of time our $p$th qubit will be in state $a_p$ and 
the rest half - in $1-a_p$. Its interaction with a separated qubit, 
say $j$th brings the deposit 
$\frac{1}{2} E_{p,j}a'_p a'_j +\frac{1}{2} E_{p,j} (1-a'_p)a'_j$ 
that is $\frac{1}{2} E_{p,j}a'_j$. 

2. Consider a different not separated qubit number $q\neq p$. 
In view of independence of time instants when NOTs are fulfiled 
on $p$th and $q$th qubits and big density $\la$ these qubits will 
be in each of states ($a_p ,\ a_q$), ($a_p ,\ 1-a_q$), ($1-a_p , \ a_q$), 
($1-a_p ,\ 1-a_q$) approximately a quarter of time. The resulting deposit 
will be $\frac{1}{4}E_{p,q} [a_p a_q +a_p (1-a_q )+(1-a_p )a_q +(1-a_p )(1-a_q )]$ $=\frac{1}{4}E_{p,q}$.

A total phase shift issued from the presence of not separated qubits 
in our procedure now is obtained by summing values from items 1 and 2 
for all $p\notin\{j,k\}$. It is 
$$\frac{1}{2}[\sum\limits_{p\notin\{j,k\}}E_{p,j}a_j +\sum\limits_{p\notin\{j,k\}}E_{p,k}a_k ]+\frac{1}{4}\sum\limits_{p,q\notin\{j,k\}} E_{p,q}.
$$
This shift can be compensated by one-qubit operations because 
the first two summands depend linearly on the qubits values 
and the second does not depend on qubits values at all. 
Thus we obtain a gate with permanent two qubits interaction 
and one-qubit operations fulfiling phase shift to $d_{j,k} a_j a_k$ 
that is required. If we take time frame $\D t$ instead of unit time 
in this procedure we obtain the phase shift to $\D t\ E_{j,k} a_j a_k$. 

Thus we can implement $U_{j,k}$ for every separated pair of qubits.

\section{Implementation of CNOT by fixed interaction}

Now show in details how to implement CNOT gate with a given pair of qubits. Let $j,k$ be fixed and omit these indexes.
Denote $\Delta E=E_1-E_2-E_3+E_4$.
If $\frac{\Delta E}{\pi} \notin Q$ 
($\frac{\Delta E}{\pi}$ is not rational number), then 
(as some physical parameters of our system, infulencing 
phases, for example, cycle period can be slighltly varied to avoid rationality 
of this parameter, the opposite case 
(rationality) can be ignored without lack of generality)
we can effectively implement  
common two-qubit gate controlled-NOT (CNOT)  
\[
CNOT=\pmatrix{1&0&0&0\cr0&1&0&0\cr0&0&0&1\cr0&0&1&0}
\]
over our pair of neighbour qubits 
by using sequence of gates only from given set 
of arbitrary one-qubit 
rotations and fixed diagonal two-qubit gate $E$ 
\[ 
E=\pmatrix{\exp{\left(i E_1\right)}&0&0&0\cr0&\exp{\left(i E_2\right)}&0&0\cr
0&0&\exp{\left(i E_3\right)}&0\cr0&0&0&\exp{\left(i E_4\right)}}
\] 
by the following way.

I. Denote gate implementing by sequential implementation of 
first qubit phase rotation A 
\[
A=\pmatrix{1&0\cr0&\exp{\left(i \left(E_1 -E_3\right)\right)}} ,
\]
second qubit phase rotation B 
\[
B=\pmatrix{\exp{\left(- i E_1\right)}&0\cr
0&\exp{\left(- i E_2\right)}} ,
\]
and gate $E$ as $U$  
\[ 
U=E \, (A \bigotimes B) = 
\pmatrix{1&0&0&0\cr0&1&0&0\cr
0&0&1&0\cr0&0&0&\exp{\left(i \Delta E\right)}} .
\] 

II. By using irrationality of $\frac{\Delta E}{\pi}$ 
it can be shown that 
\[
 \forall \varepsilon > 0 \exists m \in N \exists n \in N  
: |\Delta E n - \pi (2 m + 1)| < \varepsilon ,
\] 
i.e. for any desired accuracy $\varepsilon$ 
there exists $n=n(\varepsilon)$ so that $U^n$ will approach $\Pi$ gate 
\[
\Pi = \pmatrix{1&0&0&0\cr0&1&0&0\cr
0&0&1&0\cr0&0&0&-1} 
\] 
with given accuracy. 

III. By using relation 
\[ 
(I \bigotimes H) \Pi (I \bigotimes H) = CNOT,    
\]
where $I$ is identity matrix 
\[ 
I=\pmatrix{1&0\cr0&1} , 
\]
and $H$ is Hadamard gate 
\[
H=\frac{1}{\sqrt{2}}\pmatrix{1&1\cr1&-1} 
\]
or, in matrix form, 
\[
\frac{1}{\sqrt{2}}\pmatrix{1&1&0&0\cr1&-1&0&0\cr
0&0&1&1\cr0&0&1&-1} 
\pmatrix{1&0&0&0\cr0&1&0&0\cr
0&0&1&0\cr0&0&0&-1} 
\frac{1}{\sqrt{2}}\pmatrix{1&1&0&0\cr1&-1&0&0\cr
0&0&1&1\cr0&0&1&-1} = \pmatrix{1&0&0&0\cr0&1&0&0\cr0&0&0&1\cr0&0&1&0}
\] 
we see that controlled-NOT is finally obtained by the sequence 
\[ 
(I \bigotimes H) \left(E \, (A \bigotimes B)\right)^n (I \bigotimes H)
\]
of one-qubit rotations and gate E.

\section{Conclusion}

It is established that a quantum computer controlled by quick one-qubit transformations and with fixed permanent interaction of diagonal form between qubits is universal. It means that this type of quantum computer can implement all possible quantum algorithms by switching on and off only one-qubit gates.

\section{Acknowledgments}

We are grateful to Kamil Valiev for a stimulating discussion.

\end{document}